\documentclass[english,aps,preprint]{revtex4}
\usepackage[T1]{fontenc}
\usepackage[latin9]{inputenc}
\usepackage{array}
\usepackage{longtable}
\usepackage{graphicx}
\usepackage{amssymb}
\def \lta {\mathrel{\vcenter
     {\hbox{$<$}\nointerlineskip\hbox{$\sim$}}}}

\def\pcslash{\not{\hbox{\kern-2.1pt $p_c$}}}
\def\pgammaslash{\not{\hbox{\kern-2.1pt $p_\gamma$}}}
\def\ptslash{\not{\hbox{\kern-2.1pt $p_t$}}}
\def\Eslash{\not{\hbox{\kern-2.1pt $E$}}}

\usepackage{babel}
\let\jnfont=\rm
\def\NPB#1,{{\jnfont Nucl.\ Phys.\ B }{\bf #1},}
\def\PLB#1,{{\jnfont Phys.\ Lett.\ B }{\bf #1},}
\def\EPJC#1,{{\jnfont Euro.\ Phys.\ J.\ C }{\bf #1},}
\def\PRD#1,{{\jnfont Phys.\ Rev.\ D }{\bf #1},}
\def\PRL#1,{{\jnfont Phys.\ Rev.\ Lett.\ }{\bf #1},}
\def\MPLA#1,{{\jnfont Mod.\ Phys.\ Lett.\ A }{\bf #1},}
\def\JPG#1,{{\jnfont J.\ Phys.\ G}{\bf #1},}
\def\CTP#1,{{\jnfont Commun.\ Theor.\ Phys.\ }{\bf #1},}
\def\JHEP#1,{{\jnfont J. High \ Ener.\  Phys.}{bf #1},}
\def\RMP#1,{{\jnfont  Rev. Mod. Phys.}{bf #1},}
\def\CPL#1,{{\jnfont  Chin. Phys. Lett.}{bf #1},}

\def\Rv{\not{\hbox{\kern-1pt $R$}}}
\def\p{\not{\hbox{\kern-3pt $p$}}}

\begin{document}

\title{Single top or bottom production associated with a scalar in $\gamma p$
       collision as a probe of topcolor-assisted technicolor}

\author{Guo-Li Liu}\thanks{Email:guoliliu@zzu.edu.cn}
 \affiliation{Department of Physics, Zhengzhou University, Zhengzhou,
 China}

\begin{abstract}
In the framework of the topcolor-assisted technicolor (TC2) models,
we study the productions of a single top or bottom quark associated
with a scalar in $\gamma$-p collision, which proceed via the
subprocesses $c\gamma \to t\pi_t^0$, $c\gamma \to t h_t^0$ and
$c\gamma \to b\pi^+_t$ mediated by the anomalous top or bottom
coupling $tc\pi_t^0$, $tch_t^0$ and $bc\pi_t^+$. These productions,
while extremely suppressed in the Standard Model, are found to be
significantly enhanced in the large part of the TC2 parameter space,
especially the production via $c\gamma \to b\pi^+$ can have a cross
section of 100 fb, which may be accessible and allow for a test of
the TC2 models.
\end{abstract}

\maketitle

\section{Introduction}

The physics of the top quark \cite{review} will be intensively
studied in the coming years. The CERN Large Hadron Collider (LHC)
will copiously produce top quarks and allow to scrutinize top quark
properties. Any new physics related to the top quark will be
uncovered or stringently constrained \cite{sensitive}. One striking
property of top quark in the Standard Model (SM) is its extremely
weak flavor-changing neutral-current (FCNC) interactions due to the
GIM mechanism: they are absent at tree-level and highly suppressed
at loop-level \cite{tcvh_sm}. Therefore, the study of top quark FCNC
processes will serve as a sensitive test of the SM and a powerful
probe of new physics.

In the extensions of the SM, the top quark FCNC interactions may be
enhanced through two ways. One is that at loop-level the GIM
machanism does not work so well since the loops may contain new
particles, such as the superparticles predicted in supersymmetric
models \cite{tcvh-susy} and the mirror particles in little Higgs
models \cite{tcv-LHT}. The other is that some models like the TC2
models \cite{tc2-rev} may predict tree-level top quark FCNC Yukawa
couplings with scalar fields, which is in contrast with the SM where
the generation of fermion masses is realized by simply introducing
Yukawa couplings with only one Higgs doublet and, as a result, the
Yukawa couplings can be diagonalized simultaneously with the fermion
mass matrices. For the top quark FCNC interaction $tch$, although it
can be greatly enhanced in new physics models, the extent of
enhancement is different for different models. At the same time, in
TC2 models a large flavor mixing between the right-handed top and
charm quarks can also induce a large Yukawa coupling $bc\pi_t^+$
(this is in contrast to the usual Cabibbo-Kobayashi-Maskawa (CKM)
mixing which involves only left-handed fermions in the charged weak
current). So both the FCNC top couplings  $tch$ ($h=\pi_t^0$,
$h_t^0$) and the flavor-changing bottom coupling $bc\pi_t^+$ are
quite special in TC2 models, which may serve as a sensitive probe of
TC2 models.

Since the exotic top or bottom processes induced by the anomalous
couplings in TC2 models have been studied for hadron or linear colliders
\cite{tc-tc2}, we in this work focus on the relevant processes in
the lepton-hadron collisions.  As we know, the linac-ring type
colliders \cite{ep-review} were proposed more than thirty years ago.
Starting from the 1980's, this idea has been revisited with the
purposes of achieving high luminosities and high energies. Generally
speaking, the most popular $ep$ colliders are based on the following
suggestions. Firstly,
 THERA \cite{thera} with $\sqrt{s} =1-1.6$ TeV and $ L = 10^{31} cm^{-2} s^{-1}$
 was included in the TESLA TDR \cite{TESLA-TDR};
Secondly, the possibility to intersect CLIC ($70$ GeV) with LHC \cite{CLIC-LHC}
was discussed as a QCD explorer;
Finally, a comparison of $e$-linac and $e$-ring versions of the LHC and
VLHC based $ep$ colliders is performed in Ref. \cite{LHC-VLHC}
and the linac options are shown to be preferable.
Correspondingly, $\gamma p$ colliders \cite{rp-collision} with the same
order of luminosity and energy can be realized on the base of the
linac-ring $e p$ colliders using the Compton backscattering of laser beam off
the high energy electron beam.

It is known that the linac-ring type colliders have been playing a
crucial role in particle experiments \cite{ep-review,rp-collision}.
For example, the HERA with $\sqrt{s}=0.3$ TeV extended the
kinematical region by two orders both in high $Q^2$ and small $x$
with respect to the fixed target experiments. However, the region of
sufficiently small x ($\leq 10^{-5}$) and simultaneously high $Q^2$
($\geq 10$ GeV$^2$), where saturation of parton densities should
manifest itself, is not currently achievable. The investigation of
physics phenomena at extreme small $x$ but sufficiently high $Q^2$
is very important for understanding the nature of strong
interactions at all levels from nucleus to partons. At the same
time, the results from the linac-ring type colliders are necessary for
adequate interpretation of physics at future hadron colliders.
Concerning the on-going LHC, an $ep$ or $\gamma p$ collider with
$\sqrt{s}\simeq 1$ TeV will be very useful for the precision era of
the LHC.  Such a linac-ring collider is competitive to future
hadron or linear colliders in search for the new physics.

In this work we focus on the high energy $ep$ collider
\cite{ep-review,rp-collision} and assume the center-of-mass energy
$\sqrt{s} =1$ TeV for an illustration. We will study the productions
of a single top or bottom quark associated with a scalar in the
$\gamma$-p collision option of such an $ep$ collider, which proceed
via the subprocesses $c\gamma \to t\pi_t^0$, $c\gamma \to t h_t^0$
and $c\gamma \to b\pi^+_t$ mediated by the anomalous top coupling
$tc\pi_t^0$, $tch_t^0$ and $bc\pi_t^+$. These processes may also
serve as one of the materials to demonstrate the necessity of
constructing the $ep$ and $\gamma p$ collider. This work is
organized as follows. In sec. II we take a look at the TC2 models,
and give the Feynamn rules needed in the calculation. In Sec. III we
present the calculations for the processes at the $\gamma p$
colliders. Discussions and the conclusion are given in Sec. IV.

\section{TC2 model and the relevant couplings}

To solve the phenomenological difficulties of traditional TC theory,
TC2 theory\cite{tc2-rev} was proposed by combing TC interactions
with the topcolor interactions for the third
  generation at the scale of about 1 TeV. In TC2 theory,
  the TC interactions play a main role in breaking the
 electroweak symmetry. The Extended TC(ETC) interactions give rise to the
masses of the ordinary fermions including a very small portion of
the top quark mass, namely $\epsilon m_{t}$ with a model dependent
parameter $\epsilon \ll 1$. The topcolor interactions also make
small contributions to the EWSB, and give rise to the main part of
the top quark mass, $(1-\epsilon)m_{t}$.

After all of the dynamical symmetry breaking there are three
Nambu-Goldstone bosons (NGB) from the TC sector, and three NGBs from
top condensation sector. One linear combination of these, mostly
favoring the TC NGBs, will become the longitudinal $W_L^\pm$ and
$Z_L$.  The orthogonal linear combination will appear in the
spectrum as an isovector multiplet of pseudo-NGB(PNGB)s,
$\tilde{\pi}^a$. These objects acquire mass as a consequence of the
interference between the dynamical and ETC masses of the top quark,
i.e., the masses of the $\tilde{\pi}^a$ will be proportional to
$\epsilon$.
We refer to the $\tilde{\pi}^a$ as {\em top-pions ($\pi_t^\pm$, $\pi_t^0$)}.
For $\epsilon\lta 0.05 - 0.10$, we will find that the top-pions have
masses of order $\sim 200$ GeV. They are phenomenologically
forbidden from occurring much below $\sim 165$ GeV
\cite{Balaji:1997va} due to the absence of the decay mode $t \to
\pi_t^+ + b$.

On the other hand, the generation of a large fermion mass such as
$m_t$ is a difficult problem in  theories of dynamical EWSB. The
idea of the top quark mass as a ``constituent'', dynamical mass
generated by the presence of a condensate $\langle\bar t t\rangle$
addresses this problem, providing at the same time a source of
dynamical EWSB not requiring large amounts of new matter. In the
original top-condensation standard model~\cite{top-condensation},
the formation of the $\langle\bar t t\rangle$ condensate is fully
responsible for the masses of the SM gauge bosons as well as for the
dynamical generation of $m_t$. If the scale of the interaction
driving the condensation is $\Lambda$, then at lower energies  there
is a scalar doublet, the top-Higgs, which acquires a vacuum
expectation value (VEV). Just as in the SM, the NGB are eaten by the
$W$ and $Z$, leaving a neutral, CP-even scalar particle in the
spectrum.

In the TC2 scenario the top-pions acquire masses in the range
$m_{\pi_t}\simeq(100-300)~$ GeV. The neutral CP-even state
analogous to the $\sigma$ particle in low energy QCD, the
top-Higgs, is a $t\bar t$ bound state and its mass can be
estimated in the Nambu--Jona-Lasinio (NJL) model in the large
$N_c$ approximation, to be
\begin{equation}
m_h\simeq 2m_t\label{mthiggs}.
\end{equation}
This estimate is rather crude and it should be taken as a rough
indication of where the top-Higgs mass could be. Masses well below
the $t\bar t$ threshold are quite possible and occur in a variety of
cases~\cite{see-saw}.

 For TC2 models, the underlying interactions, i.e. topcolor
 interactions, are non-universal and therefore do not possess a
 GIM mechanism. When the non-universal
interactions are written in the mass eigenstates, it may lead to
the flavor changing coupling vertices of the new particles. Such
as, the neutral scalars predicted by this kind of models have the
flavor changing scalar coupling vertices. The coupling forms of
the scalars $\pi_t^\pm$, $ \pi_{t}^{0}$ and $ h_{t}^{0}$ to the
ordinary fermions can be written as\cite{tc2-rev,tc-tc2}:
\begin{eqnarray}
\frac{m_{t}}{\sqrt{2}F_{t}}\frac{\sqrt{v_{w}^{2}-F_{t}^{2}}}
{v_{w}}&[&K_{UR}^{tc}K_{UL}^{tt*}\bar{t}_L c_{R}h_{t}^{0}+
\sqrt{2}K_{UR}^{tt*}K_{DL}^{bb}\bar{t}_R b_{L}\pi_{t}^{+}\nonumber \\
 &+& iK_{UR}^{tc}K_{UL}^{tt*}\bar{t}_L c_{R}\pi_{t}^{0}+
\sqrt{2}K_{UR}^{tc*}K_{DL}^{bb}\bar{c}_R b_{L}\pi_{t}^{+}+h.c.],
\label{FCNH}
\end{eqnarray}
here the factor $\sqrt{v_{w}^2-F_t^2}/v_w$ ( $F_t=50$ GeV, $v_w
\simeq 174$ GeV ) reflects the effect of the mixing between the
top-pions and the would-be Goldstone bosons \cite{9702265}.
$K_{UL}$, $K_{DL}$ and $K_{UR}$ are the rotation matrices that
transform respectively the weak eigenstates of left-handed and
right-handed up-type quarks to their mass eigenstates, which can be
parametrized as \cite{tc-tc2}
\begin{equation}
K_{UL}^{tt} \simeq K_{DL}^{bb}  \simeq 1, \hspace{5mm}
K_{UR}^{tt}\simeq \frac{m_t^\prime}{m_t} = 1-\epsilon,\hspace{5mm}
K_{UR}^{tc}\leq \sqrt{1-(K_{UR}^{tt})^2}
=\sqrt{2\epsilon-\epsilon^{2}}, \label{FCSI}
\end{equation}
with $m_t^\prime$ denoting the topcolor contribution to the top
quark mass. In Eqn.(\ref{FCNH}) we neglected the mixing between up
quark and top quark.

 We can see from Eqn.(\ref{FCNH}) that only a factor $i$, the imaginary unit,
is different between the $tc\pi_t^0$ and the $tch_t^0$ couplings, so
the squares of the total amplitude of the two processes $c\gamma \to
t\pi_t^0$ and $c\gamma \to t h_t^0$ are the same with the same
scalar masses. We in following discussion, take the neutral top-pion
$\pi_t^0$ as an example unless stated otherwise.

Note that the $tc\pi_t^0$ and $bc\pi_t^+$ couplings are quite large
in the TC2 prediction, the $bc\pi_t^+$ coupling strength, $Y\sim
\frac{m_t}{F_t}\frac{\sqrt{v_{w}^{2}-F_{t}^{2}}} {v_{w}}K_{UR}^{tc}
\sim 3 K_{UR}^{tc}$ with $0.1 \leq K_{UR}^{tc} \leq 0.43$, so there
is no wonder that one expects they may induce larger contributions
to the relevant processes. At the same time, the coupling strength
yields $Y^2/4 \pi \simeq 0.11$ for $K_{UR}^{tc}=0.4$, so we can
safely conclude that the TC2 coupling as Eqn. \ref{FCNH} still makes
the perturbative expansion valid in spite of its large value.

\section{ Calculation}
\subsection{Analytical discussion}

\begin{figure}
\includegraphics[width=10cm]{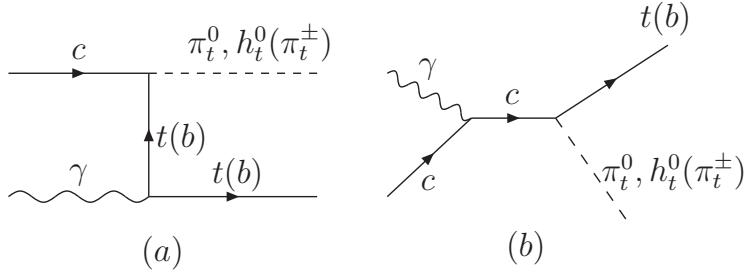}
\caption{Feynman diagrams for the tS productions $
\gamma c\to qS$ ($S=\pi_t^0$, $h_t^0$, $\pi_t^\pm$ and $q=t$ or $b$)
mediated by the anomalous couplings $tc\pi_t^0$, $tch_t^0$ and
$bc\pi_t^\pm$.\label{fig:1}}
\end{figure}

The productions of the neutral and charged top-pions at the $\gamma
p$ collision is mediated by the flavor changing $t-c-\pi_t^0$ and
$b-c-\pi_t^+$ via the subprocess $\gamma c\to t\pi_t^0$ and $\gamma
c\to b\pi_t^\pm$ with the relevant Feynman diagrams shown in Figure
1. Using the couplings given in Eqn.\ref{FCNH}, we can write the
amplitude $M_1$ and $M_2$ of the subprocesses $\gamma c\to t\pi_t^0$
and $\gamma c\to b\pi_t^+$, respectively:
\begin{eqnarray*}
&&M_{1}=eQ_c\frac{m_{t}}{\sqrt{2}F_{t}}\frac{\sqrt{v^{2}_{w}-F^{2}_{t}}}{v_{w}}K^{tc}_{UR}K^{tt*}_{UL}
\epsilon^{\mu} \bar{u_t}(P_R \frac{\pcslash+\pgammaslash}
{(p_c+p_\gamma)^2}\gamma^\mu+ \gamma^\mu
\frac{\ptslash-\pgammaslash+m_t}
{(p_t-p_\gamma)^2-m_t^2} P_R ) u_{c},\\
&&M_{2}=e\frac{m_{t}}{\sqrt{2}F_{t}}\frac{\sqrt{v^{2}_{w}-F^{2}_{t}}}{v_{w}}K^{tc}_{UR}K^{bb}_{DL}
\epsilon^{\mu} \bar{u_b}(Q_cP_R \frac{\pcslash+\pgammaslash}
{(p_c+p_\gamma)^2}\gamma^\mu+ Q_b\gamma^\mu
\frac{\ptslash-\pgammaslash+m_b}
{(p_t-p_\gamma)^2-m_b^2} P_R ) u_{c},\\
\end{eqnarray*}
Where $P_R=(1+\gamma_{5})/2$, $Q_c=2/3$, $Q_b=-1/3$, the charge of
charm quark(bottom quark) and $p_{t,c,\gamma}$ are the momentum of
top quark, charm quark and photon, respectively.

 The cross sections for the subprocess $c\gamma\to
t\pi_t^0$ and $c\gamma\to t\pi^+$ are
\begin{equation}
\hat{\sigma}(\hat s)
=\int_{\hat{t}_{min}}^{\hat{t}_{max}}\frac{1}{16\pi \hat{s}^2}
\overline M^{2}d\hat{t}\,,
 \end{equation}
with
\begin{eqnarray}
\hat{t}_{max,min} =\frac{1}{2}\left\{m_{t}^{2} +m_{\pi}^{2} -\hat{s}
\pm \sqrt{[\hat{s} -(m_{t}+m_{\pi})^{2}][\hat{s} -(m_{t}
-m_{\pi})^{2}]} \right\}.
\end{eqnarray}
Where $\sqrt{\hat{s}}$ is the center-of-mass energy of the
subprocesses in $c\gamma$ collision.

The total cross-section for the main process $\gamma q \to q\pi_t $
is obtained after the integration of $\hat{\sigma}$ over the quark
and photon distributions. For this purpose we make the following
change of variables: first expressing $\hat{s}$ as $\hat{s}=x_1x_2s$
where $\hat{s}=s_{{\gamma}q}$, $s=s_{ep}$, $ x_1=E_{\gamma}/E_e$,
$x_2=E_q/E_p$ and furthermore calling $\tau =x_1x_2$, $x_2=x$ then
one obtains $dx_1dx_2 = dx d\tau/x$. The limiting values are
$x_{1,max}=0.83$ in order to get rid of the background effects in
the Compton backscattering, particularly $e^+e^-$ pair production in
the collision of the laser with the high energy photon in the
conversion region, $x_{1,min}=0$, $x_{2,max}=1$,
$x_{2,min}=\frac{\tau}{0.83}$, $\hat{s}_{min}=(m_q+M_\pi)^2/s$. Then
we can write the total cross-section as\cite{c-gamma-distri,pdf} :
\begin{eqnarray}
\sigma(s)=\int^{0.83}_{ \hat{s}_{min}}  d\tau\int^{1}_{\tau/0.83}
dx\frac{1}{x}f_{\gamma}(\frac{\tau}{x})f_c(x)\hat{\sigma} (\hat{s})
\end{eqnarray}

where $F_{\gamma/e}$ denotes the energy spectrum of the
back-scattered photon for unpolarized initial electron and laser
photon beams given by\cite{rrcl}
\begin{eqnarray}
F_{\gamma/e}(x)&=&\frac{1}{D(\xi)} \left ( 1-x+\frac{1}{1-x}-\frac{4
x}{\xi (1-x)}+ \frac{4 x^2}{\xi^2 (1-x)^2} \right ) .
\end{eqnarray}
The definitions of parameters $\xi$, $D(\xi)$ and $x_{max}$ can be
found in \cite{rrcl}. In our numerical calculation, we choose
$\xi=4.8$, $D(\xi)=1.83$ and $x_{max}=0.83$.  $f_c(x)$ is the
distribution of charm-quarks inside the proton.



In our numerical calculation, we use the CTEQ6L \cite{cteq} parton
distribution functions and take factorization scale $Q$ and the
renormalization scale $\mu_F$ as $Q=\mu_F=m_t(m_b)+M_\pi$. To make
our predictions more realistic, we applied some kinematic cuts. We
require that the energy of $\gamma$ and the initial charm quark be
larger than $15$ GeV and the separation of two particles states be
more than $15^o$ in the center-of-mass frame. Moreover, For the
final particles, we require that the transverse momentum of each
produced particle be larger than $15$ GeV.

\subsection{Numerical results}
\label{sec:parameters}

For the SM parameters, we take $m_t=172.7$ GeV, $m_c=1$ GeV, $m_b=5$
GeV, $\alpha_e=1/128.8$ \cite{pdg} and use the one-loop running
coupling constant $\alpha_s(Q)$.

\begin{figure}
\includegraphics[width=12cm]{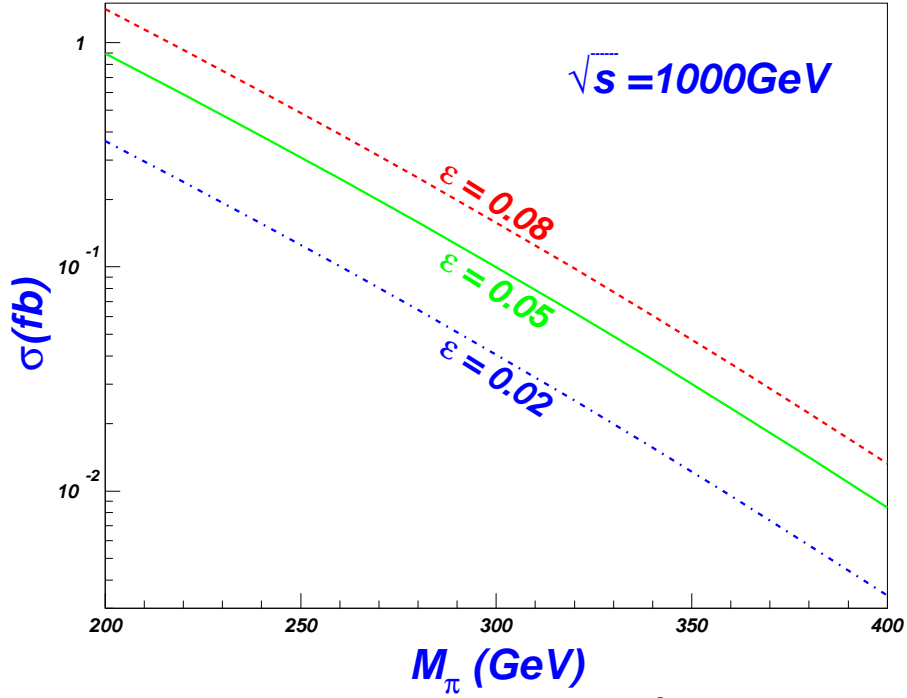}
\vspace*{-0.8cm}
\caption{The production cross section $\sigma$ of the process $c\gamma \to t\pi_t^0$
 as a function of $m_{\pi_t^0}$ for $\sqrt{s}=1$ TeV and three values of the parameter
 $\epsilon$.\label{fig:2} }
\end{figure}
\begin{figure}
\includegraphics[width=12cm]{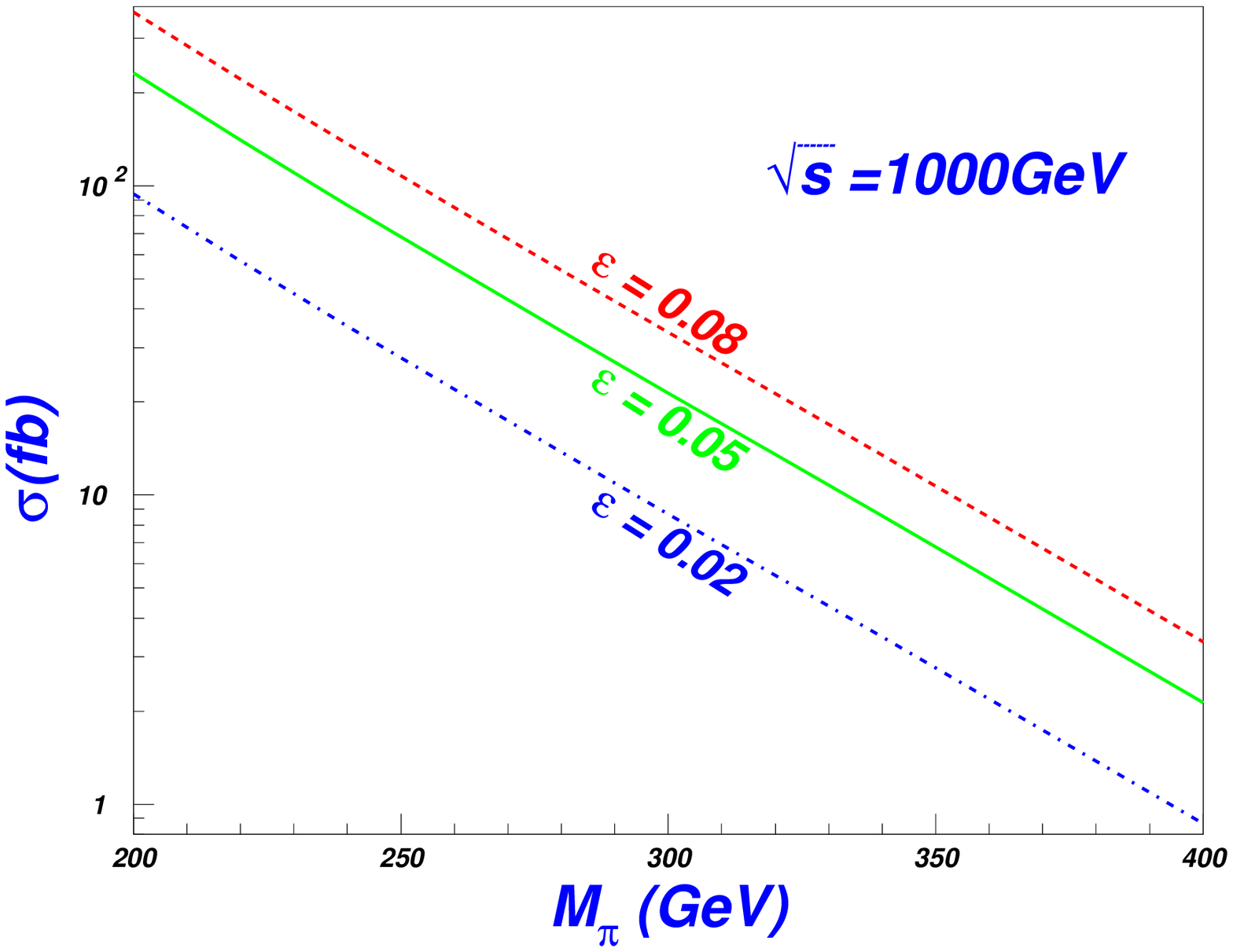}
\vspace*{-0.8cm}
\caption{The production cross section $\sigma$ of the process $c\gamma \to t\pi_t^\pm$
 as a function of $m_{\pi_t^\pm}$ for $\sqrt{s}=1$ TeV and three values of the parameter
 $\epsilon$.\label{fig:3} }
\end{figure}
\begin{figure}
\includegraphics[width=12cm]{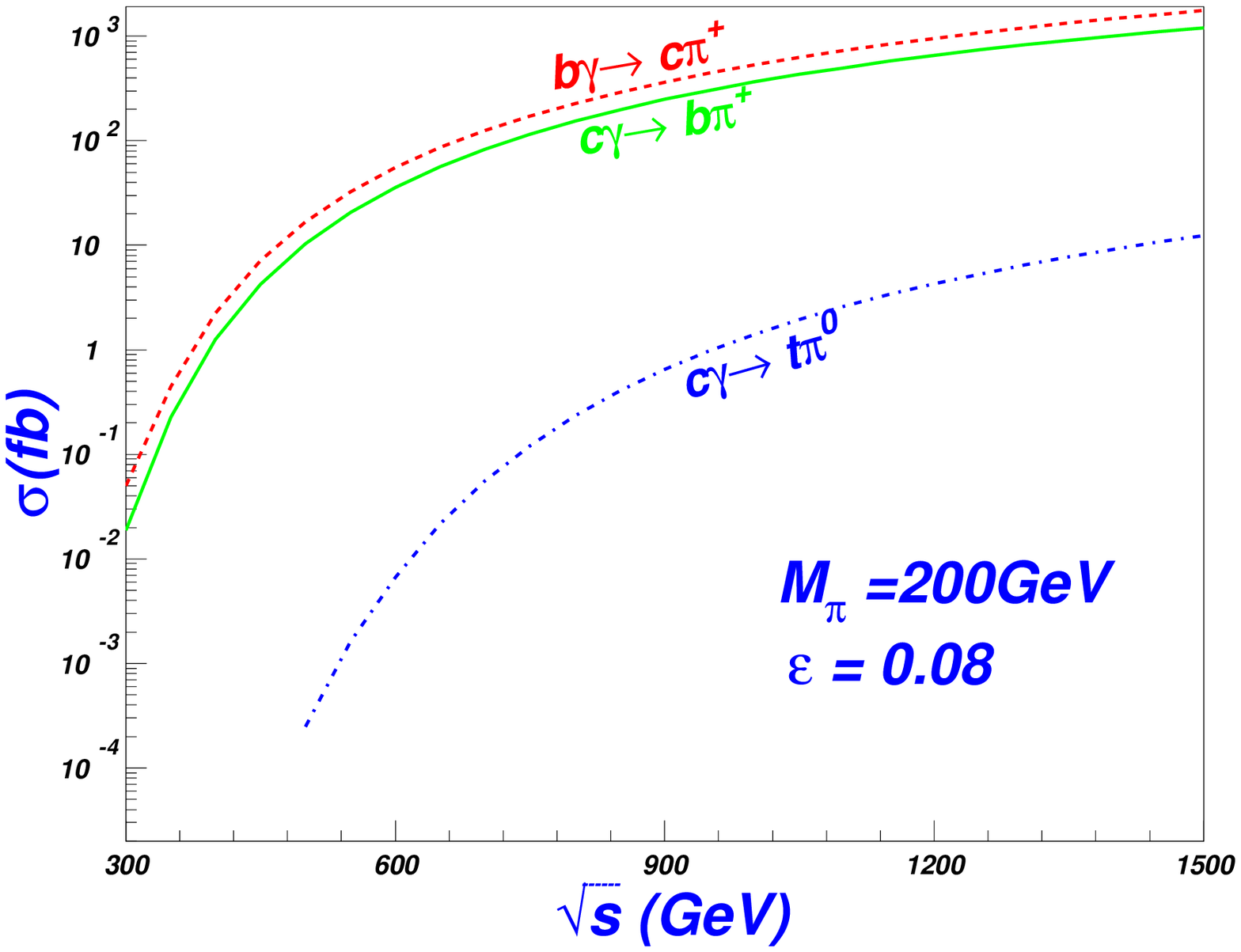}
\vspace*{-0.8cm}
\caption{The production cross section $\sigma$ of the processes
 as a function of $\sqrt{s}$ for $M_\pi=200GeV$ and $\epsilon = 0.08$.\label{fig:4} }
\end{figure}

 As for the TC2 parameters, we will consider the masses of the scalars equal to each other,
  i.e, top-pions, neutral and charged, denoted as $M_\pi$ when not considering the
difference between them. Considering the discussion in the previous
section, we assume $M_\pi$ are in the range $200-400$ GeV,
$\epsilon$ $\sim 0.01-0.08$.

The production cross sections of the top-pion $\pi^{0}_{t}$ and
$\pi^{\pm}_{t}$ at the $\gamma p$ collider are plotted in Figure 2
and Figure 3, respectively, as functions of the top-pion mass
$M_\pi$ and three values of the parameter $\epsilon$:
$\epsilon=0.02$, $0.05 $, $0.08 $ for $\sqrt{s}=1000GeV$. We can see
that the production cross sections decrease rapidly with increasing
$M_\pi$ since the final phase space are depressed by the increasing
masses of the final scalars. We can also see from the two figures
that the $\pi_t^+$ production cross section is far much larger than
that of the neutral top-pion in all of the parameter space. The
reason is that, firstly, the final phase space of the $t\pi^0_t$ is
depressed by the top mass comparing to the $b\pi^+_t$ production,
and secondly, the neutral top-pion has a large top mass propagator
which could also depress the result. For $200GeV\leq M_\pi\leq
400GeV$ and $0.02\leq \epsilon \leq 0.08$, the production cross
sections of the neutral and the charged top-pion at the $\gamma p$
collider are about $1$ fb
and $100$ fb, respectively. 
Therefore there may be hundreds of thousand $b\pi^+_t$ events to be
generated per year in most of the parameter space of the TC2 models.
Thus, it is quite easy to detect the productions of the charged
scalars via the processes $c\gamma\to b\pi_t^+$ at the $\gamma p$
collisions.

Note that the subprocess $b\gamma \to c \pi^+_t$ can also be
realized, but it is almost the same as the process $c\gamma \to b
\pi^+_t$ with the opposite fermion current. What makes the
difference between the cross section of the two processes is the
parton distribution function since one has the quark $c$ in the
initial states, while the other is the $b$ quark. We believe,
however, the difference are small, so we can safely assume the two
processes have the same cross section, which has been verified by
our calculation. To feel it, we show the dependence of the cross
section on the $\sqrt{s}$ in Fig.4, the same figure as the other
processes for comparing with each other.

 We should also note that in
Ref.\cite{0309160} the top-pion production is also discussed,
however, it is based on the effective coupling $t-c-\gamma$ at the
one-loop level, while in our discussion, the processes are induced
by the flavor changing coupling of the scalars directly
$t-c-\pi_t^0$ or $b-c-\pi^\pm_t$ at the tree level.

Despite the small probability, however, we believe that the top-pion
$\gamma p$ productions are almost free of SM backgrounds since the
$\pi^0_t t\bar c$ and $\pi^+_t b\bar c$ couplings in SM are
extremely small due to the GIM mechanism and the small $V_{bc}$
($\sim 0.04$) value in the CKM matrix.

Figure 4 displays the dependence of the cross section on the
center-of-mass (CM) energy $\sqrt{s}$ (For an simplicity of the
statement, we here assume the region of the $\sqrt s$ describes the
CM energy varying from HERA to THERA in the following discussion.),
taking $M_\pi= 200$ GeV, and $\epsilon = 0.08$, from which we can
see the cross section of charged top-pion production has to go down
as we get closer to the $m_j+M_\pi$ ($j=b,c$) threshold region and
 that all the cross sections increase with the increasing $\sqrt{s}$,
  which is one of the main goal for us to raise the collider energy,
   i. e, to
upgrade the HERA to THERA collider since the latter can provide much
larger probability of detecting the same processes. Note that the
HERA energy ($\sim 300 $ GeV) is not enough to produce the process
$c\gamma \to t\pi^0$(we plot it from $500$ GeV) since the sum of the
masses of the top quark and the top-pion boson is larger than $350$
GeV , so the THERA, with higher CM energy, can open some processes
that do not appear in the HERA collider, which is also the reason
why we carry out our calculation at the THERA, but not at the HERA
collider, i e, why we take the CM energy as $1$ TeV.

Now we further consider the signature of $b \pi_t^\pm$ production at
the $\gamma p$ collision since the rate of the $t\pi_t^0$ production
is too low. For the process $\gamma p \to b \pi^+_\pm$, $\pi_t^+$
decays to $t\bar b$ and $c\bar b$ with the branching ratio about
$70\%$ and $30\%$, respectively, with the top quark to $Wb$ and $W$
to charge lepton and the missing energy, i.e, the  $3b+l+\Eslash$
signal with $\Eslash$, the missing energy, so the mainly SM
backgrounds are $j\gamma \to jWZ$ or $jWh$, with $W\to l\Eslash$ and
$Z/h \to b\bar b$ and the jet $j$ mis-detected as $b$ quark. While
the background cross sections are very small, about $1$ fb,
Therefore if the cuts such as, to the $p_T^l$ etc., and the
b-tagging skill are employed assuming $60\%$ efficiency and $1\%$
mis-tagging, the backgrounds will be depressed violently. If we
assume only one fortieth of signal is retained (considering the top
decaying branching ration about $1/6$ to the charged lepton, the
signal may still be detected by the $\gamma p $ collider.

\section{Summary and Conclusion}
In the framework of the TC2 models, we calculated the productions of
a single top or bottom quark associated with a scalar in $\gamma$-p
collision, which proceed via the subprocesses $c\gamma \to
t\pi_t^0$, $c\gamma \to t h_t^0$ and $c\gamma \to b\pi^+_t$ mediated
by the anomalous top coupling $tc\pi_t^0$, $tch_t^0$ and
$bc\pi_t^+$. These productions, with extremely small backgrounds in
the Standard Model, were found to be significantly enhanced in the
large part of the TC2 parameter space, especially the production via
$c\gamma \to b\pi^+$ can have a cross section of 100 fb, which may
be accessible and allow for a test of the TC2 models.

\begin{acknowledgments}
We would like to thank J. J. Cao, J. M. Yang and C. P. Yuan for helpful discussion.
\end{acknowledgments}

\end{document}